\documentclass[lettersize,journal]{IEEEtran}
\usepackage{color}
\usepackage{amsmath,amsfonts}
\usepackage{array}
\usepackage{bigstrut}
\usepackage{booktabs}
\usepackage{bm}
\usepackage{textcomp}
\usepackage{stfloats}
\usepackage{url}
\usepackage{verbatim}
\usepackage{graphicx}
\usepackage{hyperref}
\usepackage{pifont}
\usepackage{algorithm}
\usepackage{algpseudocode}
\usepackage[caption=false,font=footnotesize]{subfig}
\usepackage{cite}
\def\BibTeX{{\rm B\kern-.05em{\sc i\kern-.025em b}\kern-.08em
		T\kern-.1667em\lower.7ex\hbox{E}\kern-.125emX}}
\usepackage{balance}
\usepackage{multirow}
\UseRawInputEncoding

\usepackage{graphicx}
\usepackage{float}

\begin{document}
	
\title{\huge When Vision-Language Model (VLM) Meets Beam Prediction:\\
		A Multimodal Contrastive Learning Framework}
\author{Ji Wang, \textit{Senior Member, IEEE}, Bin Tang, Jian Xiao, Qimei Cui, \textit{Senior Member, IEEE},\\
	Xingwang Li, \textit{Senior Member, IEEE}, Yingzhuang Liu, and Tony Q. S. Quek, \textit{Fellow, IEEE}
	\thanks{Ji Wang, Bin Tang, Jian Xiao are with the Department of Electronics and Information Engineering, College of Physical Science and Technology, Central China Normal University, Wuhan 430079, China (e-mail: jiwang@ccnu.edu.cn; T02052682@mails.ccnu.edu.cn; jianx@mails.ccnu.edu.cn).}
	\thanks{Qimei Cui is with National Engineering Research Center for Mobile Network Technologies, Beijing University of Posts and Telecommunications, Beijing, 100876, China (e-mail: cuiqimei@bupt.edu.cn).(\emph{Corresponding author: Qimei Cui})}
	\thanks{Xingwang Li is with the School of Physics and Electronic Information Engineering, Henan Polytechnic University, Jiaozuo 454003, China (e-mail: lixingwang@hpu.edu.cn).}
	\thanks{Yingzhuang Liu is with the School of Electronic Information and Communications, Huazhong University of Science and Technology, Wuhan 430074, China (e-mail: liuyz@hust.edu.cn).}
	\thanks{Tony Q. S. Quek is with the Singapore University of Technology and Design, Singapore 487372, and also with the Department of Electronic Engineering, Kyung Hee University, Yongin 17104, South Korea (e-mail: tonyquek@sutd.edu.sg).}	
}
\maketitle
\begin{abstract}
As the real propagation environment becomes increasingly complex and dynamic, millimeter wave beam prediction faces huge challenges. However, the powerful cross-modal representation capability of vision-language model (VLM) provides a promising approach. The traditional methods that rely on real-time channel state information (CSI) are computationally expensive and often fail to maintain accuracy in such environments. In this paper, we present a VLM-driven contrastive learning based multimodal beam prediction framework that integrates multimodal data via modality-specific encoders. To enforce cross-modal consistency, we adopt a contrastive pre-training strategy to align image and LiDAR features in the latent space. We use location information as text prompts and connect it to the text encoder to introduce language modality, which further improves cross-modal consistency. Experiments on the \emph{DeepSense-6G} dataset show that our VLM backbone provides additional semantic grounding. Compared with existing methods, the overall distance-based accuracy score (DBA-Score) of 0.9016, corresponding to 1.46\% average improvement.
\end{abstract}	
\begin{IEEEkeywords}
Vision-language model (VLM), beam prediction, multimodal, contrastive learning.
\end{IEEEkeywords}
\vspace{-5pt}		
\section{Introduction}
\IEEEPARstart The vision-language models (VLMs) have recently demonstrated the ability to align visual observations with rich linguistic priors, providing a promising approach for multimodal systems. This capability is particularly important for wireless communications, as the semantic context provided by VLMs can complement sensor data. In the millimeter wave (mmWave) networks, particularly within Vehicle-to-Infrastructure (V2I) communication systems, reliable beam prediction lays the foundation for efficient communication and improves its overall performance.~\cite{10422712},~\cite{9536953}. However, the dynamic environment and the cost of real-time channel state information (CSI) acquisition make beam prediction challenging~\cite{xiao2025wireless},~\cite{cui2025overview}. Traditionally, beam prediction methods have relied heavily on real-time CSI, but the computational overhead of acquiring it is huge, especially in rapidly changing environments. Therefore, an alternative method is needed that can predict the optimal beam without relying on real-time CSI, making the process more efficient.

Early related research focused more on using single-modal sensor data for beam prediction. Demirhan and Alkhateeb examined a radar-only scheme that reduced training overhead yet suffered from a coarse angular resolution~\cite{9771564}. Jiang \emph{et al.}\, developed a LiDAR aided approach, but its heavy preprocessing and weather sensitivity limited real-time robustness~\cite{9939167}. Morais \emph{et al.}\, explored GPS-driven methods, which performed well in open scenes, but suffered from large errors in urban scenes~\cite{10278998}. The performance of these single-sensor approaches are unstable and highly susceptible to environmental changes; consequently, researchers turned to multimodal fusion~\cite{7214350}. To overcome the instability of single modalities, Vu\v{c}kovi\'{c} \emph{et al.}\, proposed an early-fusion CNN-GRU that simultaneously ingested camera, LiDAR, RADAR and GPS streams \cite{Vuckovic2024}. These works demonstrated the promise of multimodal fusion yet revealed challenges in cross-modal alignment, which led to the contrastive learning framework proposed in this study.

Although multimodal methods have great potential, effectively aligning features from different modalities remains an important problem. Meanwhile, existing work treats position as numerical feature input. Contrastive learning aligns features of different modalities by separating different samples, and we verbalize the position information and input it into a pre-trained text encoder to further achieve cross-modal alignment.
	
In this paper, we propose a VLM-driven multimodal contrastive learning framework to enhance accuracy of beam prediction in mmWave V2I environments. The proposed model consists of three modality specific encoders. A transformer-based network extracts features from RGB images, an adaptive voxel encoder processes LiDAR, and after normalizing the GPS coordinates, a multi-layer perceptron (MLP) extracts features, while a parallel GPS-Text branch provides semantic context through language encoding, enhancing spatial understanding. The key contributions are detailed below. First, a self-supervised contrastive pre-training scheme that aligns image with LiDAR features acquired at the same timestamp is built. Features from different modalities are aligned by driving the positive pairs toward maximum similarity. Then, we develop a framework where multimodal data are encoded by specific feature encoders. Fusion occurs at the feature level, and multimodal data are complementary at this stage to improve the accuracy of mmWave beam prediction. Furthermore, we verbalize the positional information to endow the multimodal framework with VLM characteristics, thereby improving performance of beam prediction.
\section{System Architecture}
\subsection{System Model}
\begin{figure}[t]
	\centering
	\includegraphics[width=0.903\linewidth]{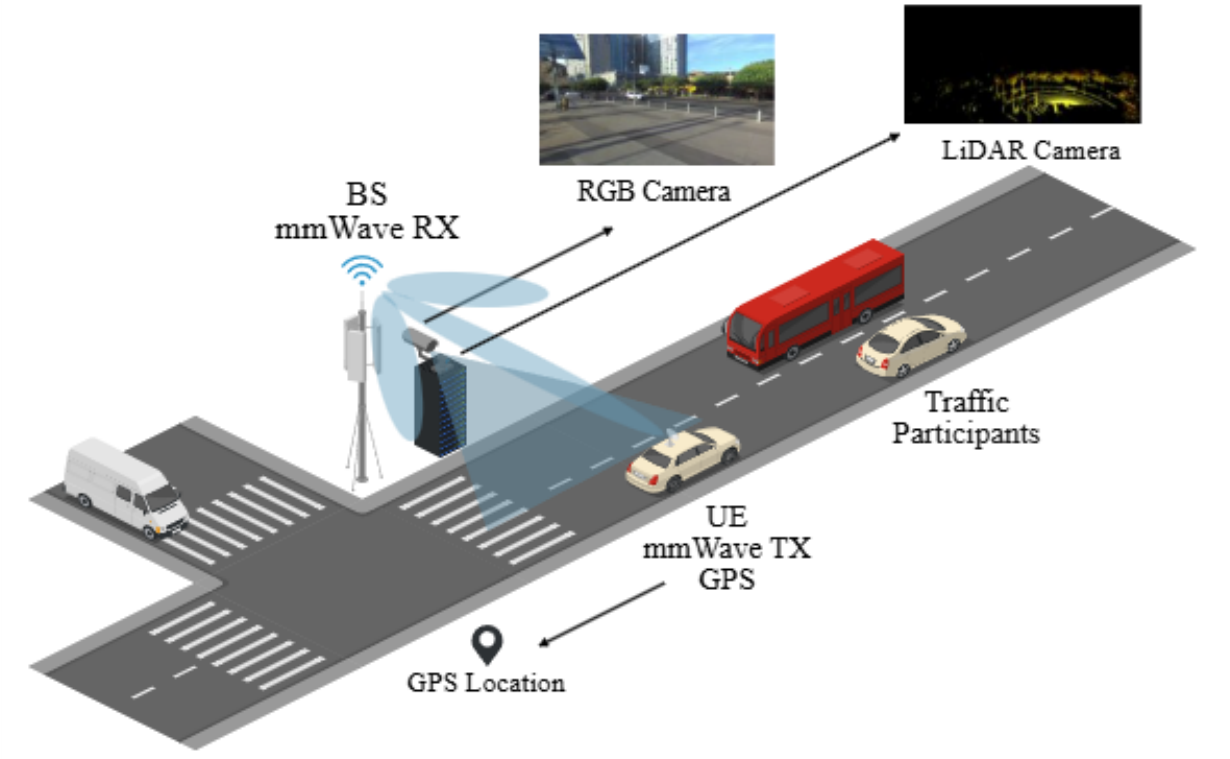}
	\vspace{-5pt}
	\caption{V2I systems with mmWave communication.}
	\label{SystemModel}
\end{figure}

Fig.~\ref{SystemModel} shows a traditional mmWave communication scenario. The communication system comprises two units: Unit~1, a base station (BS) mounted on the street; and Unit~2, a high-speed vehicle that serves as user equipment (UE). The BS is outfitted with an N-element antenna array and sensors,
using mmWave to provide communication services for UE. The UE carries a 60 GHz mmWave transmitter with an antenna, and the coordinates of UE are acquired in real time by a high-precision GPS receiver. We adopt a pre-defined beamforming codebook $\mathcal{F} = \left\{ \mathbf{f}_m \right\}_{m=1}^M$, where $\mathbf{f}_m \in \mathbb{C}^{N \times 1}$ is the $m$-th beamforming vector in the codebook and the value of $M$ represent how many vectors there are in the codebook. During the transmission process of the communication system, the channel can be defined as $\mathbf{h}_t \in
\mathbb{C}^{N \times 1}$, therefore the UE's received signal can be defined as
\begin{equation}
	y_t = \mathbf{h}_t^T \mathbf{f}_m x + z,
	\label{eq:received_signal}
\end{equation}
where $x \in \mathbb{C}$ is the complex symbol to be transmitted. The variable $z \sim \mathcal{N}_{\mathbb{C}}(0, \sigma^2)$ represents
noise at the receiver and \(\sigma^{2}\) represents the noise variance.
\vspace{-10pt}
\subsection{Problem Formulation}
The core goal of beam prediction is to avoid the high-overhead CSI acquisition process in dynamic wireless environments, and instead rely on multimodal sensors to obtain environmental perception data and select the optimal beam $\mathbf{f}_{m^*}$ from the codebook $\mathcal{F}$. The prediction problem can be formulated mathematically as
\begin{equation}
    \mathbf{f}_{m^*} = \arg\max_{\mathbf{f_m} \in \mathcal{F}} \left\| \mathbf{h}_u^T \mathbf{f_m} \right\|_2^2, 
\end{equation}
where ${m^*} \in \{1, \cdots, M\}$ denotes the index of the optimal beam. To address this challenge, we reformulate the beam selection task as a supervised classification problem, the goal is to train a predictive model $\mathcal{M}_\theta(\cdot)$, parameterized by $\theta$, that maps the input $\mathbf{x}$ to a probability distribution over all candidate beam indices as
\begin{equation}
	\mathbf{p} = \mathcal{M}_\theta(\mathbf{x}),
\end{equation}
where $\mathbf{x}=(\mathbf{x}_{\text{img}},\mathbf{x}_{\text{lidar}},\mathbf{x}_{\text{pos}})$,  it denotes the predicted beam probability distribution, and is explicitly defined as
\begin{equation}
	\mathbf{p} = [p_1, p_2, \cdots, p_M] \in \mathbb{R}^M.
\end{equation}
The final predicted beam index $\hat{m}$ is then selected as the one with the highest predicted 
\begin{equation}
	\hat{m} = \arg\max_{m \in \{1, \cdots, M\}} p_m.
\end{equation}
This formulation enables the model to learn beam selection behavior directly from labeled training data without the real-time CSI during inference. The next section outlines the architecture of our proposed framework.
\section{Proposed Multimodal Beam Prediction Framework}
\subsection{Overview of The Proposed Method}
\begin{figure}[t]
	\centering
	\includegraphics[width=1.015\linewidth]{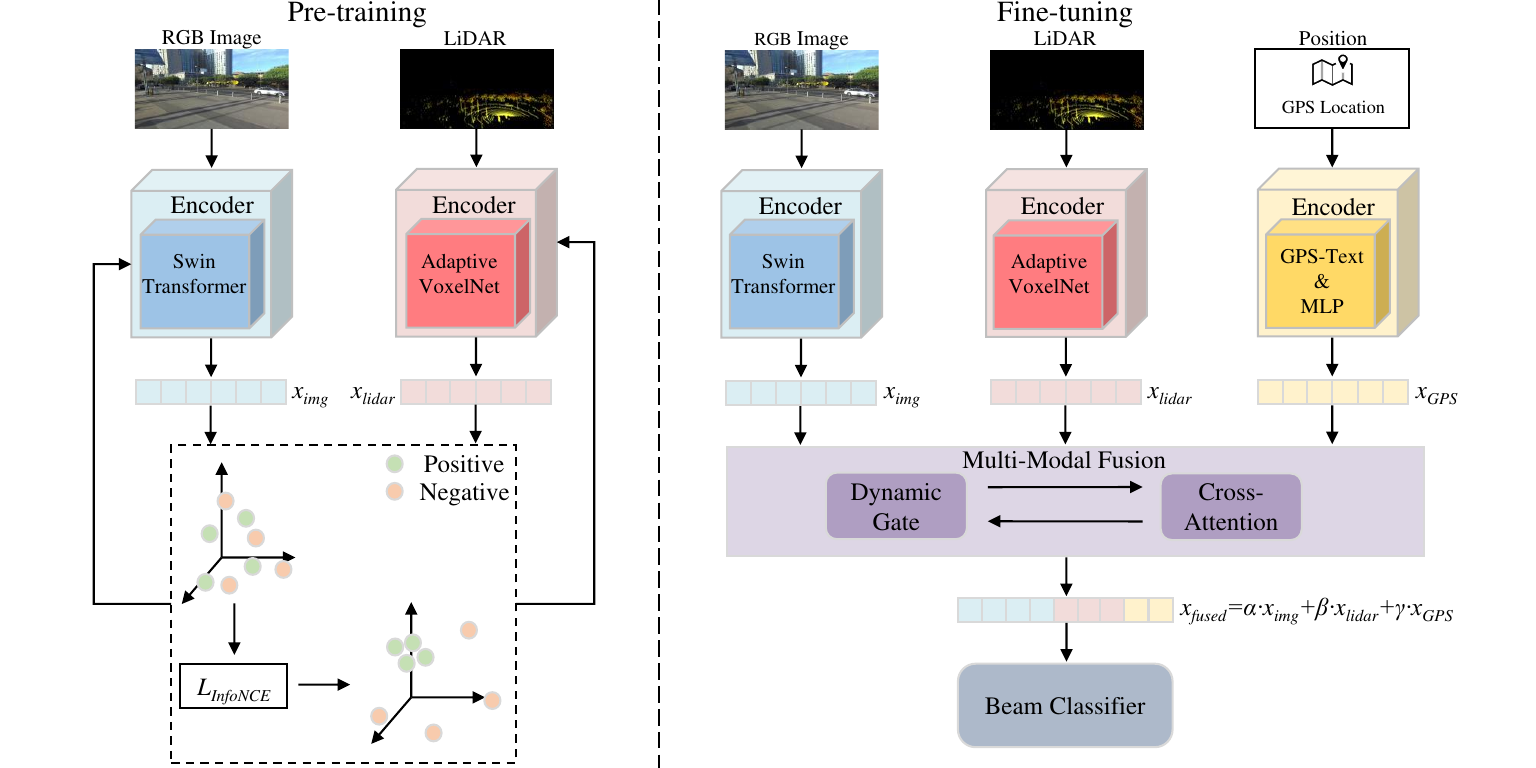}
	\vspace{-5pt}
	\caption{Overall architecture of the proposed multimodal beam prediction.}
	\label{fig:framework}
\end{figure}

Fig.~\ref{fig:framework} illustrates the overall architecture of the proposed method.
It is built around a VLM, which is composed of the following main modules: specific feature encoders, contrastive pre-training, and multimodal fusion with language branch based on GPS-Text embedding. To enhance cross-modal alignment, we introduce a contrastive learning pre-training stage, which treats image and LiDAR samples at the same time as positive pairs, thereby improving cross-modal consistency. During fine-tuning, the GPS-Text embedding is introduced and fused with the image and LiDAR features via a dynamic gating module and cross-modal attention, ensuring the language branch plays a role in the joint representation. The classifier maps the fused features into a probability vector and selects the optimal beam.
\vspace{-10pt}
\subsection{Modality-Specific Feature Encoders}
For multimodal data, each modality uses a specific feature encoder, which enables the model to extract rich features from each modality to facilitate subsequent data processing.

\textit{1) Vision Encoder:} For feature extraction of visual, we adopt a pre-trained image encoder to extract features and then map them to a shared feature space. It is processed by the vision encoder \(f_{\text{img}}(\cdot)\) and defined as
\begin{equation}
	\mathbf{x}_{\text{img}}
	= f_{\text{img}}\!\bigl(\mathbf{I};\theta_{\mathrm{v}}\bigr)
	\in \mathbb{R}^{d_{v}},
	\label{eq:vision-encoder}
\end{equation}
where \(\theta_{\mathrm{v}}\) is the set of learnable parameters of the vision branch. During vision feature extraction, the RGB image is resized and divided into non-overlapping blocks, which are linearly embedded and pass through multiple Swin blocks, and finally a feature vector is output through the local window self-attention mechanism.

\textit{2) LiDAR Encoder:} For feature extraction of LiDAR, we use an encoder based on VoxelNet. It is processed by the LiDAR encoder \(f_{\text{lidar}}(\cdot)\) and defined as
\begin{equation}
	\mathbf{x}_{\text{lidar}}
	= f_{\text{lidar}}\!\bigl(V(\mathcal{P});\theta_{\ell}\bigr)
	\in \mathbb{R}^{d_{\ell}},
	\label{eq:lidar-encoder}
\end{equation}
where \(V(\cdot)\) denotes dynamic voxelization and \(\theta_{\ell}\) is the set of learnable parameters of the LiDAR branch. Voxels are first dynamically partitioned according to the point cloud density. Within each voxel, the encoder aggregates local geometric features through voxel feature encoding (VFE) and processes them through 3D sparse convolution, and the final output is a feature vector.

\textit{3) GPS-Text (Language) Encoder:} For feature extraction of GPS, we verbalize each position into a textual prompt and feed it to a text encoder. The resulting embedding is linearly projected to the shared feature dimension to obtain the language feature $\mathbf{x}_{\text{text}}$, which defined as
\begin{equation}
	\mathbf{x}_{\text{text}} = P_{\text{text}}\!\left( \mathrm{Enc}_{\text{text}}(T_{\text{gps}});\, \boldsymbol{\theta}_{t} \right) \in \mathbb{R}^{d_{t}}.
\end{equation}
For example, we convert raw coordinates into a prompt such as: "The vehicle is currently at latitude \textit{x}, longitude \textit{y}, located to the \textit{direction} of the base station." Through this method, the GPS-Text branch introduces semantic cues, enhancing the model’s spatial understanding.
In parallel, the normalized coordinates $\hat{\mathbf{g}}$ are processed by a MLP to produce the numeric positional feature $\mathbf{x}_{\text{pos}}$, which defined as
\begin{equation}
	\mathbf{x}_{\text{pos}} = f_{\text{pos}}\!\left( \hat{\mathbf{g}};\, \boldsymbol{\theta}_{p} \right) \in \mathbb{R}^{d_{p}}
\end{equation}
We subsequently concatenate the two features and pass them through a learnable projection to obtain a unified GPS representation $x_{\text{GPS}}$, which is employed in the fusion stage.
\vspace{-10pt}
\subsection{Contrastive Pretraining Strategy}
\begin{figure}[t]
	\includegraphics[width=\linewidth]{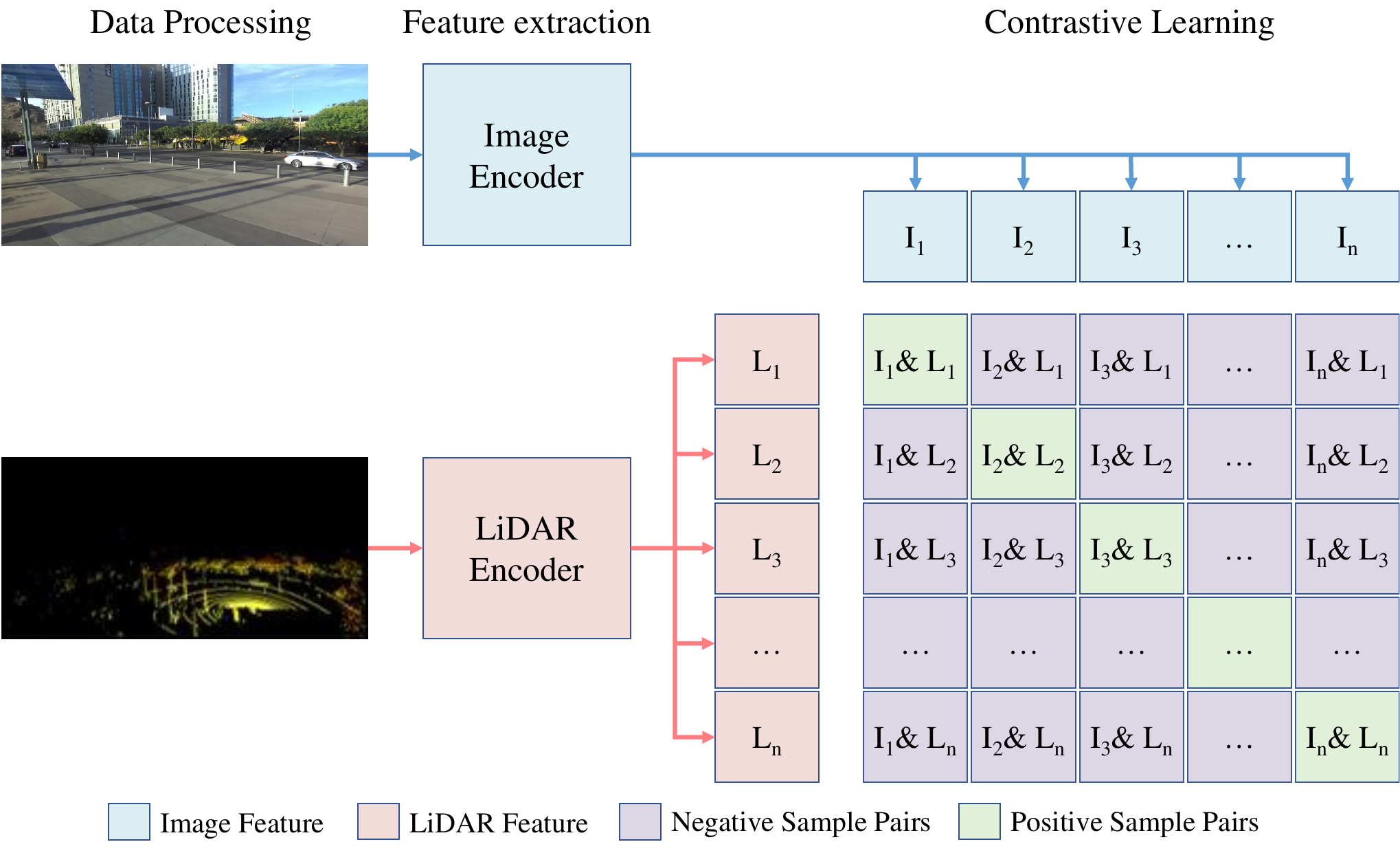}
	\vspace{-5pt}
	\caption{Illustration of the contrastive pre-training strategy.
	}
	\label{fig:contrastive}
\end{figure}

Fig.~\ref{fig:contrastive} illustrates the contrastive pre-training stage.
Since multimodal data describe the same scene from complementary perspectives, but align their features is difficult. In order to match cross-modal data at the feature level, we adopt a contrastive learning framework and pre-train the visual encoder and LiDAR encoder. In each batch, we collect N pairs of RGB-LiDAR samples synchronized in the time domain. Such a pair of images and LIDAR is considered a positive pair, and vice versa as a negative pair. After pre-training, in the latent space of feature vector mapping, the similarity between positive sample pairs can be maximized, while the similarity between negative sample pairs can be minimized.

The information noise contrastive estimation (InfoNCE) loss is adopted to align image and LiDAR embeddings in latent space. There are series of features $\left\{ \mathbf{x}^{\text{img}}_i , \mathbf{x}^{\text{lidar}}_i \right\}_{i=1}^{N}$ in a batch. The loss function can be expressed as
\begin{equation}
	\mathcal{L}_{\text{Info}} = -\frac{1}{N} \sum_{i=1}^{N} \log \frac{
		\exp\left( \text{sim}(\mathbf{x}^{\text{img}}_i, \mathbf{x}^{\text{lidar}}_i)/\tau \right)
	}{
		\sum_{j=1}^{N} \exp\left( \text{sim}(\mathbf{x}^{\text{img}}_i, \mathbf{x}^{\text{lidar}}_j)/\tau \right)
	},
\end{equation}
where $\text{sim}(\cdot,\cdot)$ denotes cosine similarity of features and $\tau$ is a temperature parameter.

This pre-training process aligns modality-specific embeddings without supervision on beam labels and improves the generalization of the encoders. The pretrained encoders are later fine-tuned jointly with the fusion modules. Compared with other alignment methods, contrastive learning provides a scalable cross-modal feature alignment method, which is well suited for real-world multi-sensor settings.
\vspace{-5pt}
\subsection{Multimodal Fusion and Beam Classification}
During fine-tuning, $\mathbf{x}_{\text{img}}$, $\mathbf{x}_{\text{lidar}}$, and the $\mathbf{x}_{\text{GPS}}$ are jointly fused by the dynamic gating and the cross-modal attention module to predict the optimal beam index.

The dynamic gating module assigns modality-specific weights based on the input, allowing the model to attend to the most informative one in varying situations, which is particularly useful in environments with sensor degradation or occlusion. Specifically, we connect the features of the three modalities and calculate the weights
\begin{equation}
	\boldsymbol{\alpha} = \text{Softmax}\left(W_g \cdot \mathbf{x} + b_g\right), 
	\label{eq:gate_weights}
\end{equation}
The weighted features are then fed into a cross-modal attention module, which captures the correlation information between different modalities and jointly represents them via semantic, geometric, and spatial information. This allows the model to down-weight degraded modalities and emphasize informative ones. The fused feature vector is defined as
\begin{equation}
	x_{\text{fused}} = \sum_{m \in \{\text{img}, \text{lidar}, \text{GPS}\}} \alpha_m \cdot \mathbf{x}_m. 
	\label{eq:fused_output}
\end{equation}

The fused feature vector is passed to a classification head consisting of a fully connected layer, which maps the feature vector to the probability distribution of all candidate beam indices through a softmax function. To optimize the model, we utilize the standard cross-entropy loss between the \begin{equation}
	\mathcal{L}_{\text{cls}} = -\sum_{m=1}^M y_m \log p_m,
\end{equation}
where $p_m$ is the predicted probability for beam index $m$, and $y_m$ is the one-hot encoded target label. This classification framework enables the model to learn an effective decision boundary, facilitating accurate beam prediction. \textbf{Algorithm \ref{alg:traandeva}} shows the specific process.
\begin{algorithm}[tbp]
	\caption{Training and Evaluation of the Proposed Model}
	\label{alg:traandeva}
	\begin{algorithmic}[1]
		\Require Input data $(\mathbf{x}_{\text{img}}, \mathbf{x}_{\text{lidar}}, \mathbf{x}_{\text{GPS}})$, beam labels ${m}$, model $\mathcal{M}_{\boldsymbol{\theta}}$, epochs $E$, learning rate $lr$
		\Ensure Predicted beam index $\hat{m}$
		\vspace{0.2em}
		\Procedure{Pretraining}{$\mathcal{M}_{\boldsymbol{\theta}}$}
		\For{$e = 1$ to $E$}
		\State Extract features $\mathbf{x}_{\text{img}}, \mathbf{x}_{\text{lidar}}$
		\State Compute InfoNCE loss $\mathcal{L}_{\text{InfoNCE}}(\mathbf{x}_{\text{img}}, \mathbf{x}_{\text{lidar}})$
		\State Update $\boldsymbol{\theta} \leftarrow \text{AdamW}(\nabla \mathcal{L}, lr, \boldsymbol{\theta})$
		\EndFor
		\EndProcedure
		\vspace{0.2em}
		\Procedure{Finetuning}{$\mathcal{M}_{\boldsymbol{\theta}}$}
		\For{$e = 1$ to $E$}
		\State Extract features $\mathbf{x}_{\text{img}}, \mathbf{x}_{\text{lidar}}, \mathbf{x}_{\text{GPS}}$ from inputs
		\State Fuse features via cross-modal attention and dynamic gating to get $\mathbf{x}_{\text{fused}}$
		\State Predict beam distribution: $\mathbf{p} = \mathcal{M}_{\boldsymbol{\theta}}(\mathbf{x})$
		\State Compute loss $\mathcal{L}_{\text{cls}}$ and update $\boldsymbol{\theta}$
		\EndFor
		\EndProcedure
		\vspace{0.2em}
		\Procedure{Testing}{$\mathcal{M}_{\boldsymbol{\theta}}$}
		\State Predict $\hat{m} = \arg\max p_m$
		\EndProcedure
	\end{algorithmic}
\end{algorithm}
\section{Numerical Results}
\subsection{Dataset Setting}
To validate the performance of the model, we use the DeepSense 6G dataset, a large-scale real-world dataset for multimodal sensing and mmWave communications~\cite{10144504}. We predict the optimal beam in scenarios 32-34. Among them, scene 32 is daytime, containing 3,235 samples; scenes 33 and 34 are nighttime, containing 3,981 and 4,439 samples respectively. In addition, the presence of pedestrians and other traffic participants increases the difficulty of the task. In the experiment, 80\% of the data is used for training, 10\% for validation, and 10\% for testing.
\vspace{-5pt}
\subsection{Evaluation Metrics}
\label{sec:eval_metrics}
The performance of our model was validated through extensive comparison studies with existing methods. To ensure that our comparative study is accurate and repeatable, we used the following evaluation indicator to evaluate the accuracy of the beam prediction task.It is distance-based accuracy score (DBA-Score)~\cite{DeepSense_Challenge}, which is defined as
\begin{equation}
	\mathrm{DBA-Score}
	= \frac{1}{K}\sum_{k=1}^{K} Y_k,
	\label{eq:dba_score_main}
\end{equation}
where $Y_k$ is defined as
\begin{equation}
	Y_k
	= 1 - \frac{1}{N}\sum_{n=1}^{N}
	\min_{1 \le k' \le k}
	\left[
	\min\!\Bigl(
	\tfrac{\lvert\hat{y}_{n,k'} - y_n\rvert}{\Delta},
	1
	\Bigr)
	\right],
	\label{eq:dba_score_yk}
\end{equation}
where $y_n$ is the ground-truth beam index, and $\hat{y}_{n,k'}$ is the $k'$-th ranked predicted index for sample $n$. DBA-Score can tolerate small beam index deviations, we use $K=3$ and $\Delta=5$ throughout our experiments.
\vspace{-5pt}
\subsection{Performance Evaluation}
\label{Evaluation}
We conducted comprehensive experiments to evaluate the proposed framework by using real V2I scenarios. All models are trained and tested using synchronized multimodal data. The evaluation uses metric defined in Section~\ref{sec:eval_metrics}: DBA-Score. The experiments are conducted on Scenario 32-34, which include both daytime and nighttime settings with varying levels of environmental complexity.

\begin{figure}[t]
	\centering
	\subfloat[Finetuning Loss Curve]{%
		\includegraphics[width=0.45\linewidth]{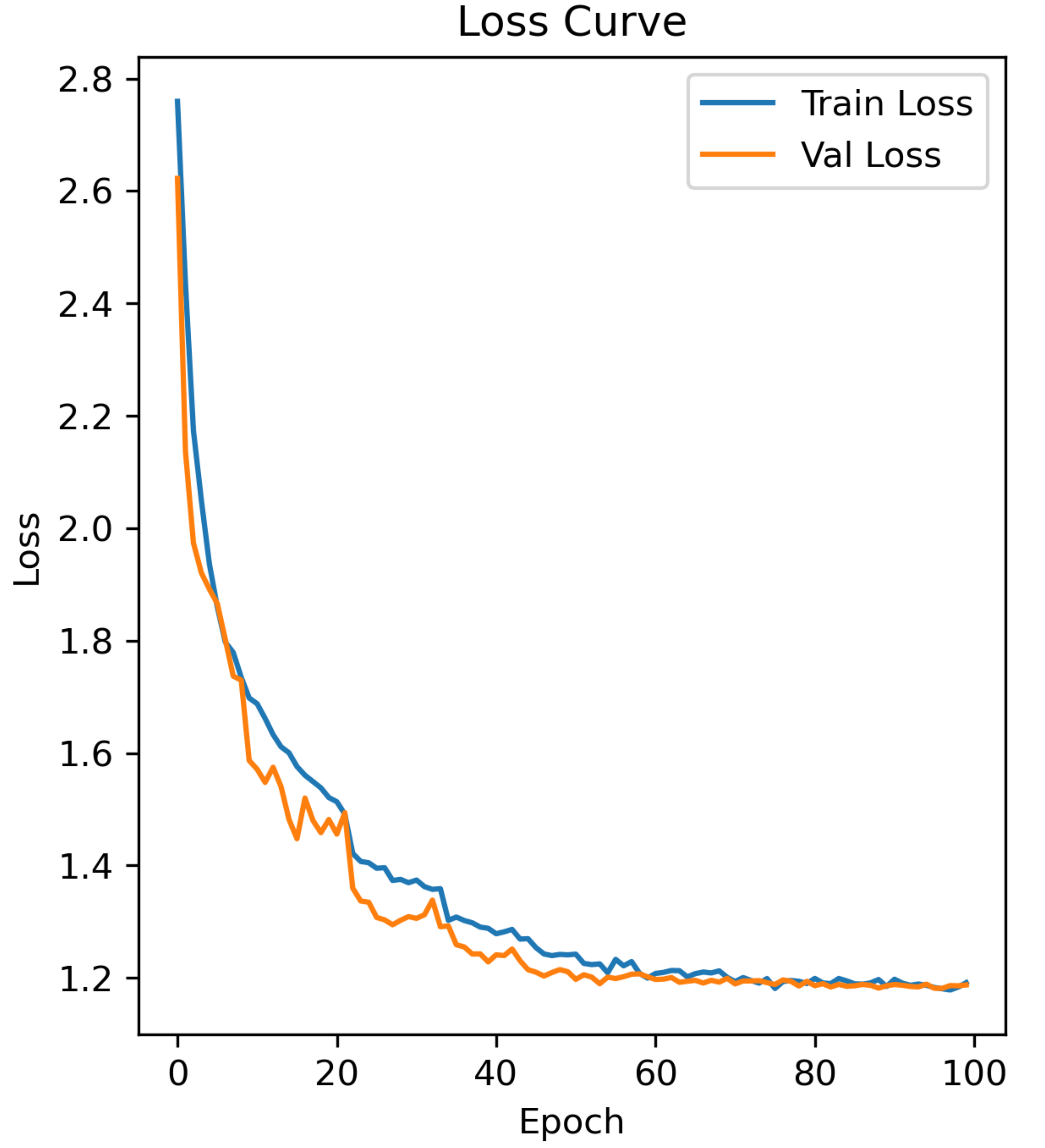}
		\label{fig:loss_curve}
	}
	\hfill
	\subfloat[Validation DBA Score]{%
		\includegraphics[width=0.45\linewidth]{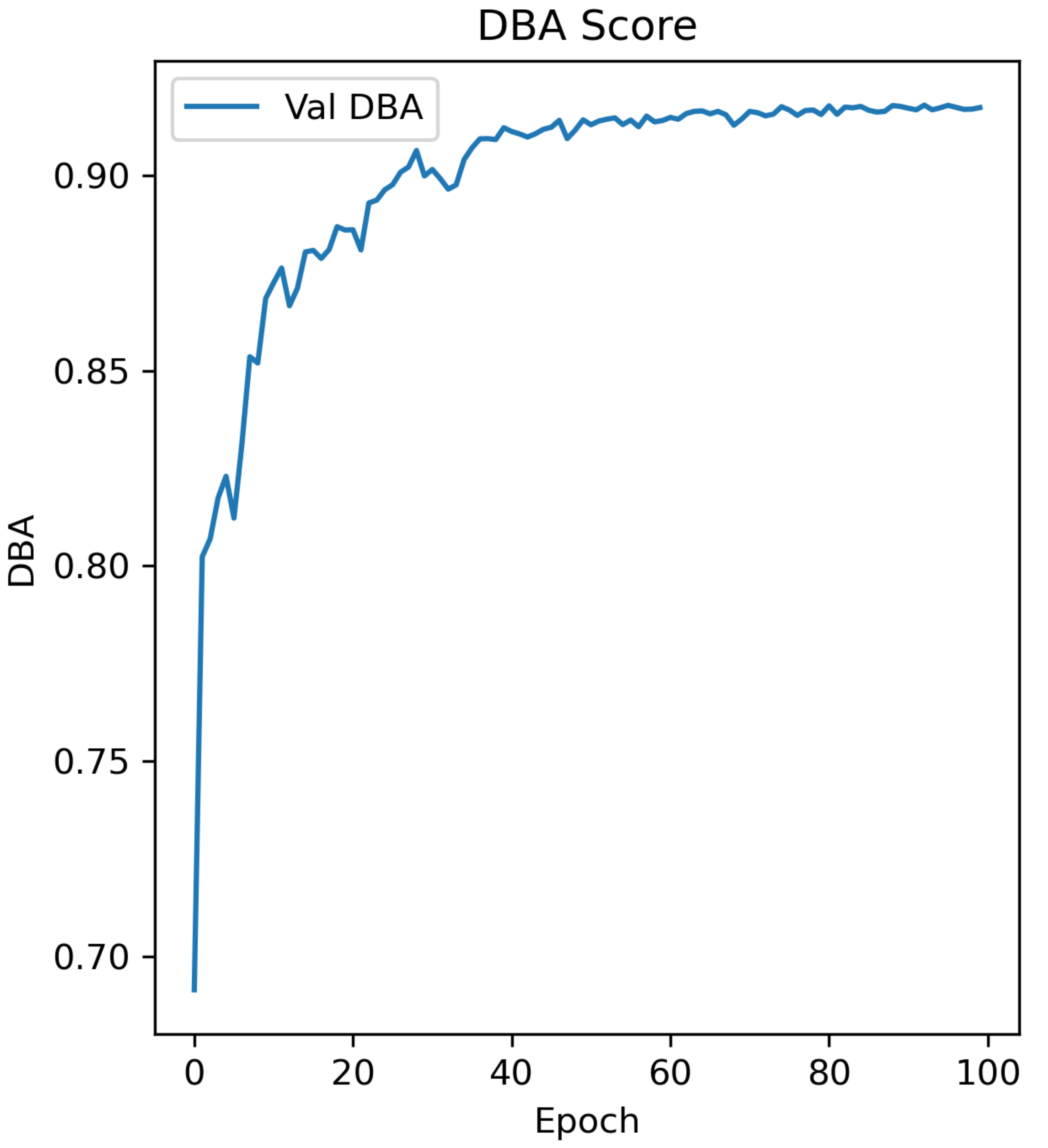}
		\label{fig:dba_curve}
	}
	\caption{Training dynamics during supervised finetuning.}
	\label{fig:finetune_curves}
\end{figure}

To further understand the training dynamics of the model, we visualized the validation DBA score and the fine-tuning loss curve. Early stopping was used to avoid overfitting caused by too many training epochs. In Fig.~\ref{fig:loss_curve}, we plot the training and validation loss over 100 finetuning epochs. The consistent downward trend and close gap between the two curves suggest stable optimization and limited overfitting. Furthermore, Fig.~\ref{fig:dba_curve} shows the DBA score on the validation set across epochs. The rapid increase and early saturation around 0.92 confirm that our model is able to perform well and converge quickly under the proposed training pipeline.
Table \ref{tab:beam_accuracy} shows that our method attains the highest DBA-Score in both daytime and nighttime scenes, which proves that our method has excellent performance in complex and diverse environments and proves the advantage of our contrastive learning pre-training method.
\begin{table}[tbp]
	\centering
	\caption{Comparison of DBA-Score and Complexity.}
	\resizebox{\linewidth}{!}{
		\begin{tabular}{lccccccc}
			\toprule
			\textbf{Method} & \textbf{S32} & \textbf{S33} & \textbf{S34} & \textbf{Overall} & \textbf{Parameters} & \textbf{FLOPs} \\
			\midrule
			Vision-Position~\cite{9771835}    & 0.8804 & 0.8778 & 0.8783 & 0.8790 & 33.02M & 7.14G \\		
			TII~\cite{tian2023multimodal}                & 0.8909 & 0.8805 & 0.8992 & 0.8904 & 66.16M & 11.42G \\
			DEQFusion~\cite{ni2023deqfusion}          & 0.8925 & 0.8768 & 0.8971 & 0.8877 & 75.61M & 11.47G \\
			AsyFFNet~\cite{9716784}           & 0.8916 & 0.8757 & 0.8979 & 0.8879 & 77.32M & 11.74G \\
			MMFF~\cite{zhang2024mmff}               & 0.8979 & 0.8804 & 0.8932 & 0.8900 & 74.04M & 56.85G \\
			ICMFE~\cite{zhu2025advancing}                   & 0.9020 & 0.8874 & 0.9074 & 0.8969 & 74.03M & 11.37G \\
			\midrule
			\textbf{Ours}                  & \textbf{0.9027} & \textbf{0.8916} & \textbf{0.9105} & \textbf{0.9016} & \textbf{75.09M} & \textbf{11.09G} \\
			\bottomrule
		\end{tabular}
	}
	\label{tab:beam_accuracy}
\end{table}
Notably, our model performs better in night-time scenes than most baselines. This is attributed to the use of LiDAR and GPS, which are more robust than RGB under poor lighting. The contrastive pretraining ensures that cross-modal alignment remains stable across varying visibility conditions.Compared with existing methods, the improvement comes not only from the integration of multimodal data, but more importantly from the proposed contrastive pre-training, which aligns the encoders of specific modalities before fine-tuning. In addition, the use of dynamic gating and cross-modal attention enables the network to suppress the modalities with degraded performance and prioritize the information-rich modalities, thus achieving consistent performance in different environments. In addition, the proposed method has 75.09M parameters and 11.09 FLOPs. These results demonstrate that our framework achieves an optimal balance between computational complexity and DBA-Score.
\vspace{-5pt}
\subsection{Ablation Experiment}
\begin{figure}[tbp]
	\includegraphics[width=\linewidth]{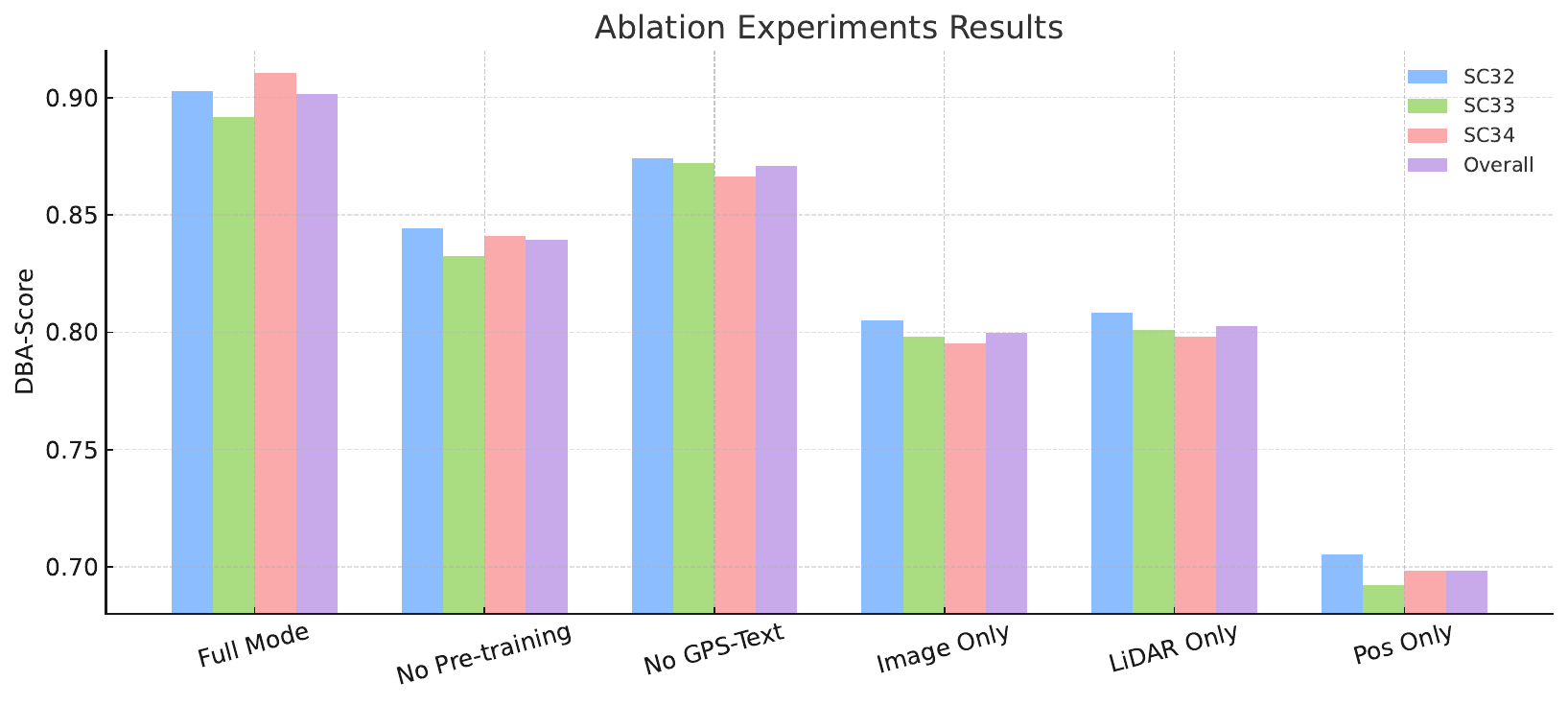}
	\vspace{-15pt}
	\caption{Results of various scenes in the ablation experiment.
	}
	\label{fig:ablation}
\end{figure}

To evaluate the individual contribution of each component, we conduct ablation  experiment using the same data splits, hyperparameters, and evaluation metrics defined in Section~\ref{Evaluation}. Several incomplete sub-models are selected from the full model by disabling each module or restricting the input to a single modality. Fig.~\ref{fig:ablation} summarizes the average results of five evaluations for each experiment.
From the above ablation experiments, we can easily find that contrastive pre-training is indispensable: once this stage is omitted, the overall DBA plummets from 0.9016 to 0.8393, confirming that the cross-modal alignment is the main driver of our gains. Eliminating the GPS-Text branch also causes a clear degradation, where the overall DBA-Score decreases from 0.9016 to 0.8709, demonstrating that verbalized positional cues provide complementary semantics. When the input is restricted to a single sensor, Image only is 0.7996, LiDAR only is 0.8025, and Position only is 0.6986, which demonstrates the importance of multimodal information for beam prediction. Ablation experiments verify the necessity of each component in the proposed method and demonstrate their synergistic effects. Removing any component will lead to a decrease in performance, which proves the effectiveness of the framework.
\vspace{-15pt}
\section{Conclusion}
This work presents a VLM-driven multimodal contrastive framework for beam prediction. The model integrates multimodal data, utilizes modality-specific encoders, and combines an attention dynamic feature fusion network. We adopt a contrastive pre-training strategy and a GPS-Text branch to align features between different modalities, thereby improving its generalization ability in complex environments. Numerical results on the real-world dataset show that our method performs better than existing baselines in both daytime and nighttime scenarios, achieving up to 0.9016 in DBA-Score, validating the effectiveness of contrastive learning pre-training and adaptive fusion. Future work will focus on enabling unseen-domain generalization and zero-shot adaptation by learning modality-invariant representations.
\vspace{-5pt}
\bibliographystyle{IEEEtran}
\bibliography{TB}

\end{document}